\documentclass[prl,twocolumn,showpacs,floatfix]{revtex4}

\usepackage{graphicx}
\usepackage{amssymb}
\usepackage[english]{babel}
\usepackage{amsmath}

\begin{document}
\title{Phase transition between synchronous and asynchronous updating algorithms}
\author{Filippo Radicchi} \author{Daniele Vilone} \author{Hildegard Meyer-Ortmanns}

\affiliation{School of Engineering and Science, International
University Bremen \footnote{Jacobs University Bremen as of spring 2007}  , P.O.Box 750561 , D-28725 Bremen , Germany.}
\noindent
\begin{abstract}
We update a one-dimensional chain of Ising spins of length $L$
with algorithms which are parameterized by the probability $p$ for
a certain site to get updated in one time step. The result of the
update event itself is determined by the energy change due to the
local change in the configuration. In this way we interpolate
between the Metropolis algorithm at zero temperature for $p$ of
the order of $1/L$ and for large $L$, and a synchronous
deterministic updating procedure for $p=1$. As function of $p$ we
observe a phase transition between the stationary states to which
the algorithm drives the system. These are non-absorbing
stationary states with antiferromagnetic domains for $p>p_c$, and
absorbing states with ferromagnetic domains  for $p\leq p_c$. This
means that above this transition the stationary states have lost
any remnants to the ferromagnetic Ising interaction. A measurement
of the critical exponents shows that this transition belongs to
the universality class of parity conservation.
\end{abstract}


\maketitle

\begin{figure}
\includegraphics[angle=0,width=0.47\textwidth,clip]{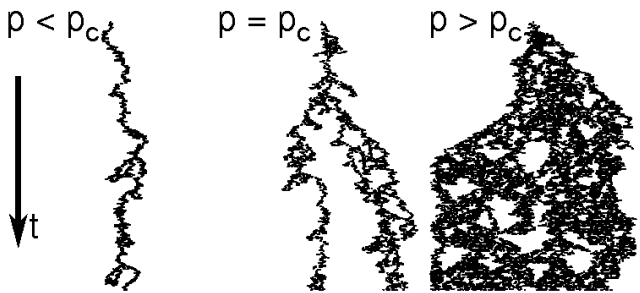}
\caption{Evolution of an isolated active bond in the subcritical,
critical and supercritical regimes. We show only the part around
the initial active bond. From left to right: $p=0.39$, $p=0.41=
p_c$, $p=0.43$.} \label{fig1}
\end{figure}

The issue of synchronous versus asynchronous updating algorithms
has  attracted much attention in connection with  Boolean
networks~\cite{kauffman:1969,kauffman:1969b,klemm-2003,klemm-2005-72,greil:2005},
neural networks~\cite{hopfield:1982,grondin:1983}, biological
networks~\cite{klemm-2005-102} and game
theory~\cite{huberman:1993,block:1999}. In a synchronous updating
scheme all the units of the system are updated at the same time.
Asynchronous updating usually means that not all units are updated
at the same time. The algorithm can be asynchronous in the sense
that each unit is updated according to its own clock, as in
distributed systems for parallel processing~\cite{ArjomandiFL83},
or it is asynchronous in the sense that only one randomly chosen
unit is updated at each step, as in Monte Carlo
algorithms~\cite{newman:1999}. In the context of thermal
equilibrium dynamics updating algorithms like the Metropolis
algorithm \cite{metropolis:1953} are designed in a way that they
drive the configurations to a set that is representative for the
Boltzmann equilibrium distribution. The dynamics of the algorithm
then enters only in an intermediate step, it is not representative
for the intrinsic dynamics of the system that is determined by the
Hamiltonian. In out-of-equilibrium systems the updating scheme may
play a more prominent role. The number of attractors in Boolean
networks, for example, increases exponentially with the system
size~\cite{samuelson:2003} for synchronous update, and with a
power for critical Boolean networks \cite{drossel:2005} for
asynchronous update. The phase diagrams of the Hopfield neural
network model~\cite{hopfield:1982,fontanari:1988} and the
Blume-Emery-Griffiths model~\cite{blume:1971,bolle:2005} depend on
the updating mode as well, while those of the Q-state Ising
model~\cite{bolle:2004} and the Sherrington-Kirkpatrick spin glass
\cite{nishi} are independent on the used scheme.
\\
Probably neither a completely synchronous nor a random
asynchronous update is realistic for natural systems. Here we
interpolate between these two extreme cases not in a more
realistic way, but in a way that allows to identify a phase
transition between the stationary states. We focus our attention
on a one-dimensional Ising model at zero temperature. We visit all
the sites and select each of them with probability $p$. Then we
update  simultaneously each of the selected candidates by flipping
their spins if the local energy of the system is not increased due
to this change. Tuning the value of $p$, we are able to
interpolate the algorithm from an asynchronous one, in the
thermodynamical limit corresponding to the Metropolis algorithm,
to a synchronous one. We observe a phase transition between the
stationary states from non-absorbing for $p > p_c$ to absorbing
ones for $p\leq p_c$, with $p_c$  the critical threshold of the
transition. The time evolution of the Ising chain can be
represented as a directed percolation (DP) process if bonds
between spins of opposite signs are called active. As our
measurements of the critical exponents show, the transition
between these stationary states belongs to the universality class
of parity conservation (PC).
\\
\vspace{0.05cm}
\\
We consider a one-dimensional lattice of length $L$. To each site
$i$, $i=1, \ldots , L$ of this chain we assign a spin variable
$\sigma_i$, where $\sigma_i$ takes the values $+1$ or $-1$. The
Hamiltonian of the system is given by $H=-J\sum_{i=1}^L \sigma_i
\sigma_{i+1}$, where $J$ is the coupling constant between
neighboring sites. Here we consider ferromagnetic couplings so
that $J=1$. Moreover, for definiteness we choose periodic boundary
conditions so that $\sigma_{L+1}=\sigma_1$. The results will not
depend on this choice.
\\
A standard local stochastic updating algorithm for studying
thermodynamic equilibrium of the Ising model is the Metropolis
algorithm~\cite{metropolis:1953,privman:1997,newman:1999}. Given a
configuration of the system $\Sigma(t)=\left\{\sigma_i(t)\right\}$
at time $t$, we pick up one site $j$ at random and flip its spin
with probability $P_j(t) = \min{\left\{1 , \exp \left[-\beta
\Delta E_j(t)\right] \right\}}$. Here $\beta=1/kT$, with $k$ the
Boltzmann constant and $T$ the temperature, while $\Delta E_j(t) =
2 \sigma_j(t) \left[\sigma_{j-1}(t)+\sigma_{j+1}(t)\right]$ is the
difference in energy that a flip of $\sigma_j$ would induce. In
particular, for zero temperature an increase in energy of the
resulting configuration is always rejected. After the single
update of the $j$-th site the time increases by $t \to t+1/L=t'$
(and $L$ single updates are considered as one time unit). The new
configuration is given by
$\Sigma\left(t'\right)=\left\{\sigma_i(t);\sigma_j(t')\right\}$,
where all sites $i \neq j$ have the same spin value as they have
at time $t$, but only the spin of the $j$-th site is eventually
flipped. It is well known that the Metropolis algorithm drives our
chain of Ising spins to the Boltzmann equilibrium distribution for
a sufficiently large number of update events.
In the context here we emphasize that the Metropolis algorithm is
fully asynchronous in the sense that we have only one spin flip
per single update event.
\\
Here we are no longer interested in the equilibrium properties of
the Ising model, but in the algorithmic dynamics applied to Ising
spins. Therefore we give up the asynchrony of the algorithm. Given
the configuration $\Sigma(t)=\left\{\sigma_i(t)\right\}$ at time
$t$, we visit all sites and select each of them with probability
$p$. The selected sites are $j_1,\ldots,j_m$ and $m$, the total
number of selected sites, is a random integer obeying the binomial
distribution $B(m,L,p) =  {L \choose m} p^m
\left(1-p\right)^{L-m}$. Each of the $m$ selected sites is then
updated according to the Metropolis rule at temperature zero, so
that the spin of the $j_v$-th site  is flipped with probability
$P_{j_v}(t)$, $\forall \; v=1,\ldots,m$. After one step of the
algorithm, the time increases as $t \to t+p=t'$, and the new
configuration is $\Sigma(t')=\left\{\sigma_i(t); \sigma_{j_1}(t');
\ldots ; \sigma_{j_m}(t')\right\}$, where all sites $i \neq
j_1,\ldots,j_m$ have the same spin value as they have at time $t$,
while the spins of the $m$ selected sites are eventually flipped.
One time unit has passed when the average number of update events
equals to the total number of sites $L$. By varying $p$ we can
``tune'' the algorithm from asynchrony, for $p$ of order of $1/L$,
to synchrony, for $p=1$. In particular, for $p$ of the order of
$1/L$ and for sufficiently large values of $L$ we recover the
usual Metropolis algorithm when two or more simultaneous
selections which occur with probability at most of the order of
$1/L^2$ become negligible. We stress the fact that the spin value
of the selected sites $j_1,\ldots,j_m$ at time $t'$ is eventually
flipped according to the actual values of the spins at time $t$,
so that they depend only on the configuration $\Sigma(t)$.
Differently from the standard Metropolis algorithm at zero
temperature, for general values of $p$, the total energy of the
new configuration $\Sigma(t')$ can be increased with respect to
the total energy of the old configuration $\Sigma(t)$. Moreover,
the long-time configurations $\Sigma(t \to \infty)$ do no longer
obey the Boltzmann distribution.
\\
In this paper, for simplicity, we focus on the case of zero
temperature. The equilibrium ground-state of the one-dimensional
Ising ferromagnet at zero temperature is one of the two
ferromagnetic states with all spins positive or negative. In
contrast, the completely synchronized dynamics does not drive the
system to the ground state, but acts as parallel algorithm and
amounts to a deterministic map $T$: $\Sigma(t+1)=T\Sigma(t)$, for
all $t$. After a transient time $t_0 \leq L/2$, the algorithm
drives the system into a cycle of length two~\cite{bolle:2004},
where the system ``flips'' between two configurations,
$\Sigma_1=\left\{^1\sigma_i\right\}$ and
$\Sigma_2=\left\{^2\sigma_i\right\}$, such that
$\Sigma_2=T\Sigma_1$ and $\Sigma_1=T\Sigma_2$, for all $t \geq
t_0$. In particular, these configurations result from each other
by an overall flip of signs in the sense that $\Sigma_2=\left\{-
\; ^1\sigma_i\right\}$ and $\Sigma_1=\left\{-\;
^2\sigma_i\right\}$. Moreover, these states are anti-ferromagnetic
because we observe domains (neighboring sites with the same value
of the spin) of a length of at most two sites. Therefore, it is
natural to study intermediate values of $p$, in particular to
focus on the transition between the ferromagnetic and the
anti-ferromagnetic configurations of the final state.
\\
Let us consider the  active bonds  of the system, where we define
a bond as active if it connects two sites with opposite spins. As
remnant of the zero-temperature Ising model only sites belonging
to at least one active bond can flip and do flip if they are
selected as candidates for the updating. Only a few elementary
processes, that involve active bonds, can take place: diffusion,
annihilation and creation. For clarity of notation, let us
indicate  as $\uparrow$ a site with positive spin and as
$\downarrow$ a site with negative  spin. Consider, for example, a
local configuration such as $\cdots \uparrow \uparrow \downarrow
\downarrow \cdots$ at time $t$:  at time $t+p$ it can evolve to
$\cdots \uparrow \uparrow \uparrow \downarrow \cdots$ or to
$\cdots \uparrow \downarrow \downarrow \downarrow \cdots$,
depending on whether they are selected for update that happens
with probability $2p(1-p)$ [diffusion], or to $\cdots \uparrow
\downarrow \uparrow \downarrow \cdots$ with probability $p^2$
[creation], or it remains unchanged with probability $(1-p)^2$.
Using the same rules, a local configuration such as $\cdots
\uparrow \uparrow \downarrow \uparrow \uparrow \cdots$ at time $t$
later, at time $t+p$, can become $\cdots \uparrow \downarrow
\uparrow \downarrow \uparrow \cdots$ with probability $p^3$
[creation], or $\cdots \uparrow \uparrow \uparrow \uparrow
\uparrow \cdots$ with probability $p(1-p)^2$ [annihilation], etc.
\ldots It is easily checked that the parity of active bonds, that
is the number of active bonds modulo $2$, is conserved.
\\
The former considerations suggest that the transition between
ferromagnetic and anti-ferromagnetic behavior (without active
bonds and with dominance of active bonds, respectively) is given
by the competition of annihilation and creation of active bonds.
In particular, the creation is favored by a synchronous updating
scheme, because a new couple of active bonds can be created only
if two neighboring sites simultaneously flip their spins.
Therefore the transition between stationary states (ferromagnetic
and anti-ferromagnetic ones) can be considered as a transition
between the asynchronous/synchronous updating schemes.
Furthermore, this transition corresponds to a $(1+1)$-dimensional
DP transition~\cite{hinrichsen-2000-49,odor-2004-76}. The active
bonds correspond to the occupied sites in DP. A qualitative
picture of this transition is shown in Figure \ref{fig1}.
We plot the time evolution of an isolated active bond (that is not
the only one in the system) for three different values of $p$. We
only display the intermediate part of the lattice around the
initial active bond. The initial configuration is chosen as
$\cdots \uparrow \uparrow \downarrow \downarrow \cdots$. In the
subcritical regime $p<p_c$, the initial active bond will remain
alone for most of the time, diffusing, and, from time to time,
creating couples of new active bonds which annihilate soon. In the
supercritical regime $p>p_c$, the average number of creations is
larger than the average number of annihilations, so that the
active bonds spread over the entire system. In the critical regime
$p=p_c$, annihilation and creation processes are balanced and the
active bonds do not spread over the system, but remain confined in
a finite region leading to the same qualitative picture as we see
in the subcritical region.
\\
In order to obtain a quantitative description of the transition,
we use as order parameter the density of active bonds $\rho$,
given by the ratio of the number of active bonds and the total
number of bonds in the lattice. The initial condition is always
chosen as $\rho(0)=1$, so that the lattice is fully occupied with
active bonds. All data points are obtained from averaging over at
least $10^3$ realizations and up to $10^5$ realizations for small
sizes of the lattice. Here, these values of $L$ are considered as
the large-volume limit in the following. Let us first determine
the critical probability $p_c$.
\begin{figure}
\includegraphics[angle=0,width=0.47\textwidth,clip]{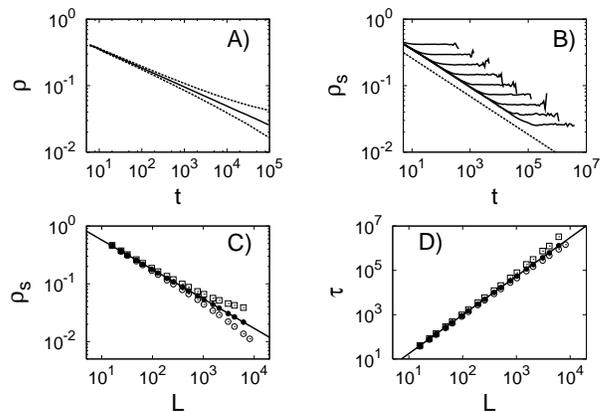}
\caption{ {\bf A)} Time decay of the density of active
 bonds $\rho$ in the subcritical ($p=0.40$),
critical ($p=0.41 = p_c$) and supercritical ($p=0.42$) regime,
from bottom to top, respectively. {\bf B)} Average density of
active bonds $\rho_s$ surviving up to time $t$ over samples,
plotted as a function of time at the critical point for $L=16, 32,
64, 128, 256, 512, 1024, 2048, 4096$ and $6144$, from top to
bottom respectively. The dotted line has a slope of
$-\delta=-0.286(1)$. {\bf C)} Average density of active sites over
the surviving samples $\rho_s$ as a function of size $L$ of the
lattice for $p=0.40$ (empty dots) $0.41$ (full dots) and $0.42$
(empty squares). The full line has slope
$-\beta/\nu_{\perp}=-0.51(1)$. {\bf D)} Average relaxation time
$\tau$ as a function of the size $L$ of the lattice for $p=0.40$,
$0.41$ and $0.42$ [the symbols are the same as in C)]. The full
line has slope $z=\nu_{\parallel}/\nu_{\perp}=1.746(2)$.}
\label{fig2}
\end{figure}
In Figure \ref{fig2}A) we plot the time behavior of $\rho$ for
three different values of $p$ and $L=10^4$. As we can see, for
$p=0.41$ $\rho(t)\sim t^{-\delta}$ with $\delta=0.286(1)$, while
for $p=0.40$ $\rho(t)$ decreases with negative curvature, for
$p=0.42$ $\rho(t)$ it increases with positive curvature. The
positive curvature characterizes the different phase. Therefore,
within the given accuracy, we locate the critical threshold $p_c$
as the largest value such that the curvature is non-positive. In
this way we obtain $p_c=0.41(1)$ as the critical point.
Determining $p_c$ with higher precision would amount to an
increase in the linear size of the lattice $L$ and the time for
observing the positive curvature of $\rho$. Here we do not perform
this kind of computationally expensive simulations, but calculate
the critical
exponents without increasing the precision in $p_c$.\\
{\bf Critical exponents.} Along the determination of $p_c$ we have
already read off the exponent $\delta$ from the time evolution of
$\rho$: for $p<p_c$ and in the thermodynamic limit $L\to\infty$,
we should observe $\rho(t) \sim t^{-\alpha(p)}$, with $\alpha(p)$
a continuous and monotonic decreasing function of $p$ and enclosed
by curves with $\alpha(p)=1/2$ for $p \to 0^+$, as in the case of
the standard Metropolis algorithm~\cite{privman:1997,newman:1999},
and $\alpha(p_c)=\delta$, as observed in our numerical simulation.
Therefore, the negative curvature observed in the subcritical
regime, is a finite-size effect. Also in the supercritical regime
the curve bends down to zero as a finite-size effect after a
sufficiently large time, but the positive curvature signals the
onset of the new (supercritical) phase.\\
{\bf Finite-size scaling analysis.} Finite-size effects in the
density of active bonds are manifest in two ways: if we follow the
time evolution of a certain configuration of a chain of length
$L$, we observe a power-law decay of active bonds up to a certain
time $t_d(L,p)$ for $p\leq p_c$. After that, either the density
drops to zero faster than a power-law, this happens for most
configurations, for which a fluctuation drives the system into the
absorbing state, or, in the minority of evolutions, the number of
active bonds fluctuates around a plateau, before the plateau drops
to zero in the end. Configurations of this minority are called
{\it surviving} configurations up to time $\tau$. Now it is easier
to locate the onset of the plateau than the onset of a faster
decay, and therefore to study the finite-size scaling of the
density of active bonds $\rho_s$ of surviving configurations as a
function of $L$, averaged only over the surviving
realizations~\cite{jensen-1994}. In Figure \ref{fig2}B) we plot
the time behavior of $\rho_s$ at the critical point for several
values of the size $L$. As we can see, after an initial transient
in which $\rho_s$ decreases as $t^{-\delta}$, it reaches a
stationary value depending on $L$. This value vanishes in the
thermodynamical limit, since the plateau is a finite-size effect.
We average along the values of the plateau (since averaging over
time and over different realizations are assumed to be
equivalent), for different system sizes. In Figure \ref{fig2}C)
this average value is plotted as a function of $L$ for three
values of $p$ ($p=0.40$, $p=0.41=p_c$ and $p=0.42$). Again, at the
critical point we find a power law decay $\rho_s\sim
L^{-\beta/\nu_{\perp}}$, with $\beta/\nu_{\perp} = 0.51(1)$, while
the decay deviates from the power-law behavior in both the
subcritical and the supercritical regimes. (Here the exponent
$\beta$ characterizes the behavior of the order parameter
$\bar{\rho}(p) \sim \vert p_c-p\vert^\beta$ , $\nu_{\perp}$ the
spatial correlation length $\xi(p)\sim \vert p_c-p\vert
^{-\nu_{\perp}}$ both in the vicinity of $p_c$.)\\
Moreover, from the finite-size scaling analysis we calculate the
dynamical exponent $z=\nu_{\parallel}/\nu_{\perp}$, where
$\nu_{\parallel}$ characterizes the time-like correlation length
$\tau(p)\sim \vert p_c-p\vert^{-\nu_{\parallel}}$. The exponent
$z$ is derived from the relaxation time $\tau$ needed by a finite
system to reach the absorbing configuration [$\rho \left( t \geq
\tau \right)=0$]. In Figure \ref{fig2}D) $\tau$ is plotted as a
function of $L$ for $p=0.40$, $p=0.41=p_c$ and $p=0.42$. At
$p=p_c$ we find again a power-law dependence $\tau\sim L^z$, with
$z=\nu_{\parallel}/\nu_{\perp}= 1.746(2)$, while for different
values of $p$ $\tau$ behaves differently from a power-law
behavior, as it is seen in Figure \ref{fig2}D). Furthermore, in
Figure \ref{fig3} we verify the finite-size scaling relation
$\rho\left(p_c,L,t\right) \sim L^{-\beta/\nu_{\perp}}
f\left(t/L^z\right)$, where $f(\cdot)$ is a suitable universal
function.
\begin{figure}[hbt]
\includegraphics[angle=0,width=0.47\textwidth,clip]{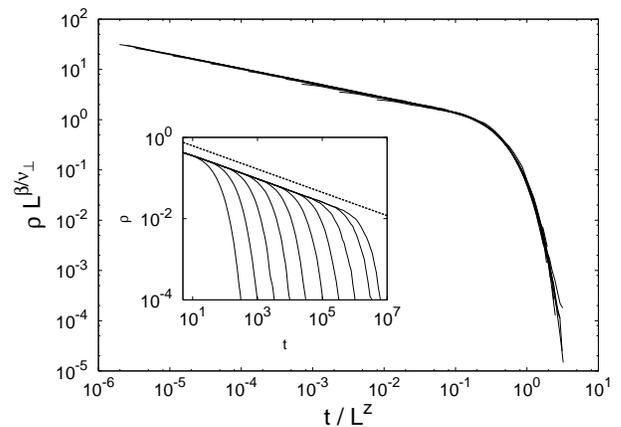}
\caption{The main plot shows the finite-size scaling of the
density of active bonds $\rho$. The inset shows the unscaled data,
where the dotted line has a slope equal to $-\delta$. The datasets
are the same as in Figure \ref{fig2}. } \label{fig3}
\end{figure}
Finally, we determine the static exponent $\gamma$ from the growth
of fluctuations in the order-parameter susceptibility, defined as
$\chi = L (\langle\rho^2\rangle - \langle\rho\rangle^2)$, via
$\chi_s\sim L^{\gamma/\nu_{\perp}}$ at $p_c$ (where the index $s$
again refers to an average over the surviving configurations). We
find $\gamma \simeq 0$, but we do not report any figure here. We
further determine the dynamical exponent $\eta$ and check the
exponents $\delta $ and $z$ in the following way. Starting from a
configuration like $\cdots \uparrow \uparrow \downarrow \uparrow
\uparrow \cdots$, that is a ferromagnetic configuration with only
two active bonds, we numerically compute the survival probability
$P(t)$  (that is the probability that the system had not entered
the absorbing state up to time $t$), the average number of active
bonds $\bar{n}(t)$ and the average mean square distance of
spreading $\bar{R^2}(t)$ from an arbitrary selected site. These
quantities are  expected to behave at the critical point $p_c$ as
$P(t) \sim t^{-\delta}$, $\bar{n}(t) \sim t^{\eta}$ and
$\bar{R^2}(t) \sim t^{2/z}$
\cite{hinrichsen-2000-49,odor-2004-76,jensen-1994}. {}From these
simulations we find $\eta \simeq 0$ and values of $\delta$ and $z$
consistent with the former ones within the error bars. The figures
are not displayed.
\\
\\
In conclusion, we have calculated the static exponents
$\beta/\nu_{\perp}=0.51(1)$ and  $\gamma \simeq 0$ and the
dynamical exponents $\delta=0.286(1)$,
$\nu_{\parallel}/\nu_{\perp}=z=1.746(2)$ and $\eta \simeq 0$.
These values are consistent with the ones conjectured for the PC
universality class $\beta/\nu_{\perp} = 1/2$ , $\gamma=0$, $\delta
= 2/7$, $\nu_{\parallel}/\nu_{\perp} = 21/12$ and
$\eta=0$~\cite{hinrichsen-2000-49,odor-2004-76,jensen-1994,
jensen-1994b,benavraham-1994-50,zhong-1995-28,janseen:1997,menyhard-1996-29,menyhard-2000-30}.
This is expected because periodic boundary conditions for Ising
spins as well as the updating rules we used preserve the parity of
the number of active bonds, and for free boundary conditions only
the boundaries may violate the conservation of parity which leads
to a negligible effect. Even if parity is non-conserved as in case
of the non-equilibrium kinetic Ising model
~\cite{menyhard-1996-29,menyhard-2000-30}, for which spin-flip and
spin-exchange  dynamics are mixed, the PC universality class is
observed. An increase in the dimensionality of the lattice would
allow a mean-field description of the transition, since the PC
universality class has the critical dimension
$d_c=2$~\cite{hinrichsen-2000-49,odor-2004-76}. Our preliminary
numerical simulations for this case show qualitatively similar
phases as in $d=1$, but with the transition point shifted towards
$1$. Our work represents the first systematic interpolation
between a synchronous and asynchronous updating scheme. The
existence of a phase transition sheds some light on the
interpretation of stationary states whenever they depend on the
updating mode. Beyond the transition these states may have lost
any remnants of the "intrinsic" dynamics (in our case the
ferromagnetic Ising interaction). Instead, they are representative
for the dynamics of the updating algorithm itself.


\end{document}